\documentclass[pra,twocolumn,longbibliography]{revtex4-2}

\usepackage{braket}
\usepackage[T1]{fontenc}
\usepackage{amsmath}
\usepackage{amsthm}
\usepackage{amssymb}
\usepackage{amsfonts}
\usepackage{mathtools}
\usepackage{dsfont}
\usepackage[normalem]{ulem}
\usepackage{graphicx}
\usepackage{siunitx}
\usepackage[version=4]{mhchem}
\usepackage{datetime}
\usepackage{enumitem}   
\usepackage{color}
\usepackage{newtxtext,newtxmath}
\usepackage{microtype}
\usepackage{braket}
\usepackage{physics}
\usepackage{xr}
\usepackage{ragged2e}
\usepackage{natbib}
\usepackage{hyperref}
\usepackage{multirow}
\usepackage{caption}
\usepackage{subcaption}
\usepackage{booktabs}
\usepackage{enumitem}
\newlist{steps}{enumerate}{1}
\setlist[steps, 1]{label = Step \arabic*:}

\newcommand{\probP}{\text{I\kern-0.15em P}}

\theoremstyle{definition}

\renewcommand{\i}{\mathrm{i}}

\begin{document}

\newcommand{\figdir}{figures}
\newcommand{\liege}{Institut de Physique Nucléaire, Atomique et de Spectroscopie, CESAM, Universit\'e de Li\`ege, 4000 Liège, Belgium}
\title{Deep-learning-based continuous attacks on QKD protocols}

\author{Théo Lejeune}
\affiliation{\liege}
\author{François Damanet}
\affiliation{\liege}

\begin{abstract}
The most important characteristic of a Quantum Key Distribution (QKD) protocol is its security against third-party attacks, and the potential countermeasures available. While new types of attacks are regularly developed in the literature, they rarely involve the use of weak continuous measurement and more specifically machine learning to infer the qubit states. In this paper, we design a new individual attack scheme called \textit{Deep-learning-based continuous attack} (DLCA) that exploits continuous measurement together with the powerful pattern recognition capacities of deep recurrent neural networks. As a minimal model, we present its performances when applied in the case of the BB84 protocol with intrinsic noise in the communication channel. Our results suggest that our attack's performances lie between the ones of standard intercept-and-resend attacks and of the optimal individual attack, namely the phase-covariant quantum cloner. Our attack scheme demonstrates deep-learning-enhanced quantum state tomography applied to QKD, and could be generalized in many different ways, notably in the cases of quantum hacking attacks targeting implementation vulnerabilities that could compromise the security of QKD protocols.

\end{abstract}
\date{\today}
\maketitle

\section{Introduction}

In the last decades, the demand for fast, secure, and reliable data connections has significantly increased. To meet this demand, it is essential to enhance the computational power of network systems through high-performance technologies. 
Quantum computing is one such technology, showing a potential to outperform current classical computing systems.
Quantum computing-assisted communications have therefore been extensively studied and developed in recent years, and hold great promise for improving communications and security in today's networks~\cite{QC_networks, QKD_selector}.
    
At the same time, quantum computing also represents a threat in terms of security, in particular related to some asymmetric cryptographic algorithms such as RSA (Rivest-Shamir-Adleman)~\cite{RSA}, a public key cryptosystem still used in many secure data transmissions to this day. Indeed, standard encryption techniques such as RSA could be broken through Shor's algorithm, a quantum algorithm factoring large integers exponentially faster than the best-known classical algorithms~\cite{Shor_algo}. 

While current quantum computing technology is still far from being enough advanced to break RSA, this motivated the elaboration of encryption techniques based on quantum mechanical properties. Quantum Key Distribution (QKD), which aim is to implement the exchange of a secure private key over a public insecure channel between two parties, is the most famous category of quantum cryptography protocols~\cite{renner_QKD_security}. The first and most known QKD protocol is the BB84, proposed by Charles Bennett and Gilles Brassard in 1984~\cite{bb84}, which uses linearly polarized photons traveling in an optical fiber. Among others are the B92 that uses entangled particles~\cite{B92}, the Differential-phase-shift which does not require a basis selection~\cite{DPS} or the Decoy State protocol designed to overcome photon number splitting attacks~\cite{Decoy_state}. 

One of the main motivations to look for quantum cryptography protocols over classical ones originates from the perturbative nature of measurement in quantum mechanics. Indeed, any spy acting on a communication channel will influence the states of the qubits used to store the private key bits traveling inside, because of the collapse of the wavefunction, which makes the spy more easily detectable than in classical communication protocols. In other words, any attempt to retrieve information from the system will inevitably introduce some disturbance, described by the Information-Disturbance theorem~\cite{Miyadera_2007, Fuchs_1996}.

Amongst the most studied types of attacks are the \textit{Intercept-and-Resend} type, \textit{Photon Number Splitting} (PNS)~\cite{PNS_attack} and \textit{Trojan Horse} (also called Large Pulse attack)~\cite{Trojan_horse, Large_pulse}. New Intercept-and-Resend attacks have been developed recently, such as \textit{Blinding}~\cite{Blinding}, \textit{Time shift}~\cite{Time_shift, time_shift_practical} or \textit{Dead-time}~\cite{Dead_time}. While these attacks fall under the category of \textit{quantum hacking}—exploiting vulnerabilities in practical implementations—many involve projective measurements that typically disturb the quantum state of the photon. For a comprehensive review, we refer the reader to Xu. \textit{et al.}~\cite{2020_Xu}. If the two parties, say Alice and Bob, do not deploy specific countermeasures against the attacks~\cite{QKD_counter}, they can however usually find a way to decide or not on the presence of a spy on the quantum communication channel~\cite{QKD_entropy_accumulation, attack_ratings_QKD, BB84_review, QKD_model}, by sacrificing a few bits of the sifted key (i.e., by sharing their measurement results on a public channel) and calculating the Quantum Bit Error Rate (QBER), which is defined as the rate of incorrect results Bob gets when measuring the qubits sent by Alice in the right basis~\cite{QBER}, i.e.,
\begin{equation}\label{eqQBER}
    \mathrm{QBER} = \frac{N_{\mathrm{error}}}{N_{\mathrm{total}}},
\end{equation}
where $N_{\mathrm{total}}$ is the total number of qubits received where Bob used the right measurement basis, and $N_{\mathrm{error}}$ is the number of incorrect results he gets among these qubits. In the case of a perfect quantum communication channel, the QBER should be zero. However, in the presence of a spy, the qubits states are usually altered and the QBER non-zero despite Alice and Bob using the same basis for the qubits, which should thus signal the two parties that something went wrong. Complications then arise because in practice, the quantum channel is not perfectly isolated from its environment (e.g., the optical fiber could be leaky~\cite{Kozubov2019,Miroshnichenko2012,Miroshnichenko2012a}) and Bob measurement apparatus could be defective, contributing to another cause of QBER enhancement. Hence, distinguishing an attack from intrinsic noise in the channel is not always easy. Despite this, several fundamental papers established the unconditional security of the BB84 protocol in the early 2000s~\cite{Lo_1999, 2000_Shor_Preskill, renner_QKD_security, Devetak_2005, Koashi_2009}. The security is unconditional in the sense that no assumption is made on Eve's attacks: Eve can perform any measurement scheme on the channel and the channel may be subject to dissipation, Alice and Bob will upper-bound Eve's obtainable information and, provided the QBER is below a certain threshold, reduce it to an arbitrarily low level using privacy amplification.

The recent development of Machine Learning (ML) and Deep Learning (DL) techniques has led to many improvements in QKD, whether to enhance existing protocols or to detect attacks more easily. Indeed, DL has been used to identify if an attacker is present or not in an IoT network, based on the final key length~\cite{2021_Al_Mohammed}. In Continuous-Variable QKD (CV-QKD), ML has been used for \textit{wavelength-attack} recognition~\cite{2020_He} and \textit{calibration-attack} recognition~\cite{2020_Mao_calibration_attack_defense}. A single neural network was trained to detect \textit{calibration-attacks}, \textit{LO-intensity-attacks} and \textit{saturation-attacks}, or two types of hybrid attack strategies~\cite{2020_Mao_general_ML_defense}.
Tunc \textit{et al}. implemented a recurrent neural network and a support vector machines algorithm to protect the BB84 protocol against attacks~\cite{2023_Tunc}. Such tools have also been implemented in CV-QKD protocols for, among other things, noise filtering~\cite{2022_Zhang}, wavefront correction~\cite{2023_Long}, state classification~\cite{2018_Li_b}, parameter estimation~\cite{2018_Liu} and parameter optimization~\cite{2019_Su}.

However, only a couple of works have investigated how artificial intelligence could be used to develop more effective attacks: a convolutional neural network was trained to help the eavesdropper choose the best opportunity to launch an \textit{entanglement-distillation-attack}~\cite{2019_Huang}, an ML algorithm was shown to be able to analyze the power originating from the integrated electrical control circuit to perform a \textit{power-analysis-attack}~\cite{2021_Zheng}, and a quantum circuit implementation of the BB84 protocol was interpreted as a quantum machine learning task, allowing to find a cloning algorithm outperforming known ones~\cite{2024_Decker}. Also, a deep convolutional neural network was used to monitor the electromagnetic emissions of a QKD emitter (\textit{Deep-learning-based radio-frequency side-channel attack}~\cite{DL_RF_attack}).
Despite neural networks demonstrating a better ability to perform quantum state reconstruction using partial information and fewer measurements than classical state tomography~\cite{Torlai_2018, Quek_2021, Gray_2018, Gebhart_2023}, the literature does not show extensive research on ML/DL-based attacks. Often, Eve is modeled as introducing an ancillary quantum system (i.e., a probe) that interacts unitarily with the traveling qubits through some channel, before either being measured directly (individual attacks) or stored in order to perform a coherent or collective measurement later. The power of the attack is evaluated by an information-theoretic upper bound, such as the mutual information between the probe system and Alice's bits which represents the maximum information Eve could, in principle, extract, regardless of the specific measurement.

In this paper, we develop a new type of individual attack based on continuous measurement~\cite{Wiseman} of single polarized photons and apply it for concreteness in the context of the BB84 protocol, by contrast to other works that usually apply this kind of measurement on CVQKD. The general motivation is to evaluate the performance of a specific measurement scheme that theoretically produces only a small perturbation to the qubits, by contrast with the effects of projective measurement usually involved in the other types of attacks. In particular, we investigate how the effects of the spy measurement can be optimally hidden by the intrinsic noise of the quantum communication channel by minimizing the increase in QBER due to the measurement. At the same time, we investigate how neural networks can effectively use the information extracted by these types of measurement, to infer a more significant part of the key than conventional means.
To do so, we feed the outcome of the continuous measurement, also called homodyne photo currents, to a Long Short-Term Memory (LSTM) recurrent neural network~\cite{LSTM} to retrieve the initial states of the photons sent by Alice, which compose the sifted key generated. The main point of this paper is not to question the security of the BB84 protocol, but rather to show an application of deep learning in QKD, namely deep-learning-assisted quantum state tomography, and to raise awareness on its potential use in the context of attacks.
 
This paper is organized as follows: In Sec.~\ref{model} we first summarize the BB84 protocol, present our model of the qubit dynamics in the quantum communication channel when subject to intrinsic dissipation and continuous measurement, investigate how a spy could use the outcome of this measurement to obtain the initial state of the qubit and present the neural network we implemented to do so.
In Sec.~\ref{results} we compare the results obtained via a basic projective measurement (Intercept-and-Resend attack) and our measurement scheme. In Sec.~\ref{info_gain_key_rate}., we discuss our attack scheme in terms of information gain and how it fits into the thresholds established by the information-disturbance principle, and calculate the typical key rate Alice and Bob should achieve to secure the protocol. In particular, we compare our attack performances against the ones of an optimal individual attack strategy for the BB84 protocol: the covariant-phase quantum cloner~\cite{Bruss_2000, Fuchs_1997}.
Finally, in Sec.~\ref{conclu}, we conclude and discuss potential perspectives of our work.

\section{Model and methods}\label{model}

In this section, we first remind how the standard BB84 protocol works briefly, before presenting how we model the dynamics of the qubits used in the protocol when they are subject to dissipation and continuous measurement in the quantum communication channel. Then, we describe the neural network that we envision a spy could use to retrieve the states of the qubits based on the continuous measurement they performed in the channel. Finally, we present how we quantify the impact of the measurement on the protocol.

\subsection{BB84 protocol} \label{BB84_steps}
The BB84 protocol, sketched in Fig.~\ref{summary_intro}, implements a shared secret key between two parties by storing private key bits in linearly polarized states of photons. There are four initial states: vertically and horizontally polarized states represented by $\ket{0}$ and $\ket{1}$ respectively, and two diagonally polarized states defined as
\begin{equation}
    \begin{aligned}
        \ket{+} &= \frac{\ket{0} + \ket{1}}{\sqrt{2}},  \\
        \ket{-} &= \frac{\ket{0} - \ket{1}}{\sqrt{2}}.  
    \end{aligned}
\end{equation}
These states define the Pauli-Z eigenbasis $\{\ket{0}$,$\ket{1}\}$ and the Pauli-X eigenbasis $\{\ket{+}$,$\ket{-}\}$~\cite{QCS}. The protocol can then be summarized as follows (see Fig.~\ref{summary_intro})~\cite{Nielsen_Chuang}:
 \begin{steps}
        \item Alice chooses a random data bit string $b$ (e.g., $b = 01011\dots)$. She encodes each data bit randomly as the quantum states $\ket{0}$ or $\ket{+}$ if the corresponding bit of $b$ is 1 and $\ket{1}$ or $\ket{-}$ if the corresponding bit of $b$ is 0.
                    
        \item Alice sends the resulting qubits to Bob via an optical fiber.
                    
        \item Bob receives the qubits and measures each of them in the Pauli-X or Pauli-Z eigenbasis at random.
                    
        \item Via the public channel Alice and Bob compare, for each qubit, the basis chosen by Alice to encode it and the basis chosen by Bob to measure this same qubit.
        They discard all the qubits where the two bases do not correspond.

        \item Alice selects a subset of her bits to check on the interference caused by the spy -- the so-called Eve --, and tells Bob which bits she chose. They both announce and compare the values of the check bits via the public channel and calculate the QBER given by Eq.~(\ref{eqQBER}). If it is higher than a threshold (typically $11\%$~\cite{2000_Shor_Preskill}), they abort the protocol.
    \end{steps}
    Alice and Bob now each possess a \textit{sifted key}, which may slightly be different because of the dissipation and spy-induced QBER. 
    To increase the security of the protocol, two additional steps are performed. The first is information reconciliation (also called error correction) to correct bits that have been modified by dissipation and spying~\cite{Nielsen_Chuang, info_reconciliation}. The second is privacy amplification, which consists in passing the generated key in a hash function to exponentially decrease Eve information.~\cite{Nielsen_Chuang,privacy_ampl,privacy_ampl2}. By doing so, Alice and Bob obtain a shorter but more secure final key. It is important to note that hash functions decrease the key length, such that the spy looses information in the data compression process.
    This last step requires, however, an upper bound estimate of Eve’s information on the corrected key. Thus the bigger the part of the sifted key Eve has, the more bits Alice and Bob must sacrifice in the process.
    
\subsection{Dissipative qubit dynamics conditioned on measurement}

  \begin{figure}
        \centering
        \includegraphics[width = 0.975\linewidth]{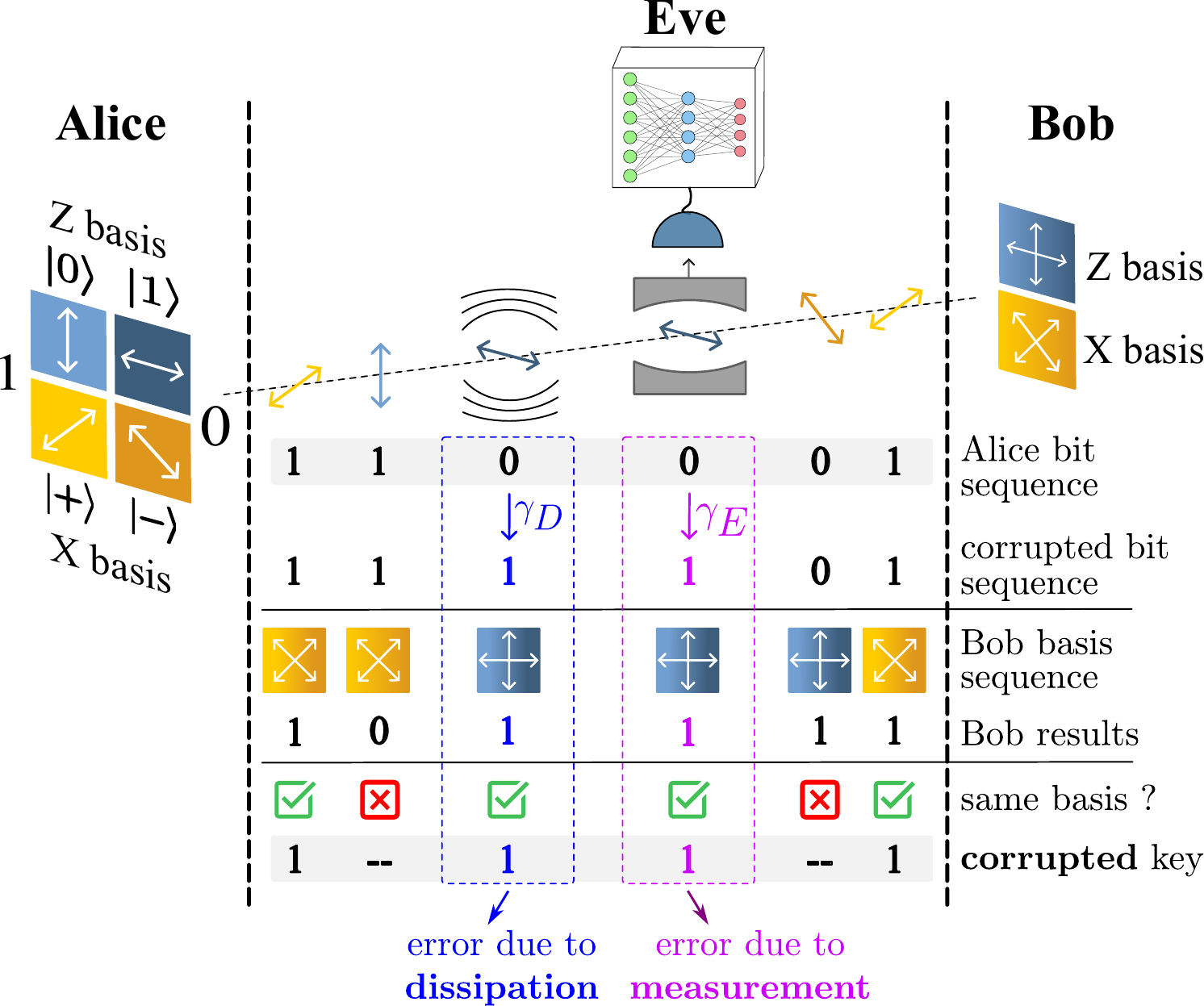}
        \caption{
        \justifying{Sketch representation of our attack scheme applied on the BB84 protocol. Alice sends linearly polarized photons to Bob via an optical fiber, while traveling they are subject to intrinsic dissipation in the fiber with a rate $\gamma_D$ and to weak measurement with a rate $\gamma_E$ performed by Eve on a certain portion of the fiber, the output of which being treated by a neural network to determine the initial state. In the sketched example, Alice sends six photons to Bob, among which three are polarized in the Pauli-X eigenbasis and three in the Pauli-Z eigenbasis. Bob measures randomly in one of these two bases each of the photons received, and the basis chosen is the right one for four of these. However, even if the measurement basis is right, there are two errors. The sketch highlights the two possible mechanisms for errors: intrinsic dissipation (blue dashed square) and measurement by a third party (purple dashed square).
        In this example, if Alice and Bob compare all the measurements when they chose the same basis, the QBER [Eq.~(\ref{eqQBER})] would be $50\%$.}}
        \label{summary_intro}
    \end{figure}

For concreteness, we model the dynamics of each individual qubit in the quantum communication channel as subjected to i) intrinsic dissipation acting on the channel and ii) a continuous measurement performed by a spy (see Fig.~\ref{summary_intro}). More specifically, we consider the following stochastic master equation (written here in Itô form)~\cite{1993_Wiseman}
\begin{equation}
\begin{aligned}
        d\rho_J= &-i\left[H,\rho_J\right]dt + \gamma_D\mathcal{D}[d]\rho_Jdt \\
        &+ \gamma_E\mathcal{D}[e]\rho_Jdt + \sqrt{\gamma_E\eta}\mathcal{H}[e]\rho_J dW,
        \label{SME}
\end{aligned}
\end{equation}
where $H$ is the channel Hamiltonian defined as $H=\omega\sigma_z$ for the initial states $\ket{0}$ or $\ket{1}$ and $H=\omega\sigma_x$ for $\ket{+}$ or $\ket{-}$~\cite{Kozubov2019} with $\sigma_x = |0\rangle\langle 1| + |1\rangle \langle 0|$ and $\sigma_z = |0\rangle\langle 0| - |1\rangle \langle 1|$ the standard Pauli operators, where $\rho_J$ is the density operator of the qubit conditioned on the measurement with efficiency $\eta \in [0,1]$ of the homodyne current~\cite{Wiseman}
\begin{equation} \label{Jhom}
J dt = \sqrt{\eta\gamma_E} \langle e + e^\dagger \rangle dt + dW,
\end{equation}
and the superoperators $\mathcal{D}[o]$ and $\mathcal{H}[o]$ are defined as
\begin{align} \label{mathcalD}
   &\mathcal{D}[o]\boldsymbol{\cdot} = o \boldsymbol{\cdot} o^\dagger - \frac{1}{2}(o^\dagger o \boldsymbol{\cdot} - \boldsymbol{\cdot} o^\dagger o) \\
   &\mathcal{H}[o]\boldsymbol{\cdot} = o\boldsymbol{\cdot} + \boldsymbol{\cdot} o^\dag - \mathrm{Tr}\left[o\boldsymbol{\cdot} + \boldsymbol{\cdot} o^\dag\right]\boldsymbol{\cdot},
        \label{mathcalH}  
\end{align}
for a given operator $o$.
In Eq.~(\ref{SME}), the first line represents the effect of the unitary dynamics of the channel governed by the Hamiltonian $H$ as well as the effect of the intrinsic dissipation produced by the operator $d$ occuring at rate $\gamma_D$, while the second line represents the effect of the eavesdropping produced by the operator $e$ at rate $\gamma_e$, which decomposes into an incoherent term and a non-linear stochastic term, where $dW$ is a Wiener increment satisfying E$[dW]=0$ and $dW^2 = dt$. For concreteness, we set throughout this work the dissipation operator to be
\begin{equation}
    d = \sigma_x,
\end{equation} to model the dissipation as a bit-flip error, but any other choice could be made without any additional complexity (see Sec.~\ref{sec_key_rate}), depending e.g. on the specific open system model considered for the optical fiber.

In terms of practical implementations of the continuous measurement, while we do not intend in this paper to provide a specific detailed scheme, we foresee that this could be realized indirectly, for examples, via the monitoring of an auxiliary field that couples to the photons on a certain portion of the optical fiber, or by extracting a small amplitude of the signal via a low-ratio beam splitter or a directional coupler. In Appendix~\ref{adia}, we show how to derive Eq.~(\ref{SME}) from the homodyne detection of such an ancillary field and its adiabatic elimination.

The measured current~(\ref{Jhom}) allows in principle the spy to estimate the state of the qubit from the expectation value of $\langle e + e^\dagger \rangle$, as explained in the next section. The goal of the spy consists in i) minimizing the impact of their measurement on the quantum channel and ii) retrieving at best the initial qubit state sent by Alice.

\subsection{Standard quantum state tomography}

Since the spy wants to obtain the initial state of the photons from a continuous measurement, the data he has access to is the homodyne photo current of each photon he measured. Since the initial state is random, the spy cannot estimate the state from averaging over many photo currents: they have to estimate it from a single photo current for each qubit. In this scenario, using standard quantum state tomography (QST), which is the process of reconstructing the quantum state of a system from repeated measurements of a set of observables, is very difficult. 

In~\cite{Impossibility_measure}, D'Ariano and Yuen reviewed a variety of concrete measurement
schemes~\cite{Inhibition_msrt, meaning_wave_fct, reversible_msrt, QND_reversible, Reversible_msrt_spin}, and concluded that it is practically impossible to determine the wave function of a system from a single copy of it. More recent works on tomography, including plain averaging or maximum likelihood methods~\cite{homodyne_tomography}, direct inversion, distance minimization, maximum likelihood estimate with radial priors and Bayesian mean estimate~\cite{tomography_methods}, or Bayesian Homodyne and Heterodyne tomography~\cite{Bayesian_tomography}, also show that it is difficult to reconstruct efficiently the initial state from one copy of the system or one measurement. In fact, without the measurement of a complete set of observables (a quorum), there is not enough information for the reconstruction as different states may give the exact same statistics on an incomplete set of observables~\cite{high_dim_homodyne_tomo,photon_tomography}. Hence, it is inefficient to use standard quantum state tomography techniques to estimate a qubit state from a single homodyne measurement on a photon, which motivated us to employ a deep learning approach, as explained below.

\subsection{Neural network quantum tomography based on the measurement}

\begin{figure*}[ht!]
    \centering
    \includegraphics[width = 0.8\linewidth]{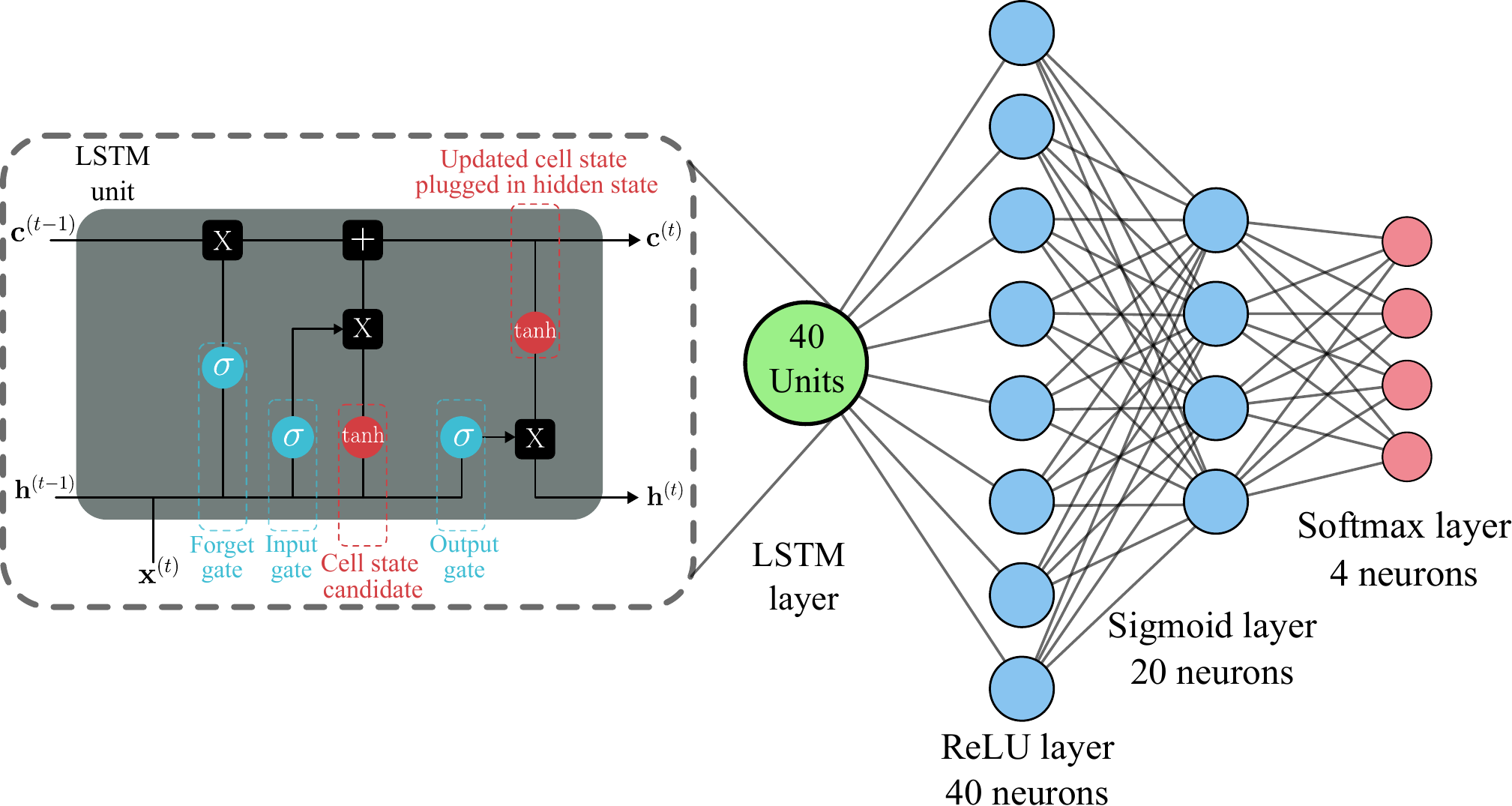}
    \caption{\justifying{Sketch representation of the LSTM architecture used in this paper. The input layer is composed of $100$ LSTM units, and is followed by two dense hidden layers of $40$ ReLU neurons and $4$ linear neurons respectively. On the left is a representation of one LSTM recurrent unit, composed of three gates with sigmoid activation functions (forget, input and output). These 3 gates determine which information from the prior hidden state must be erased, taken into account and stored respectively.}}
    \label{model_illu}
\end{figure*}

The homodyne photo currents resulting from the measurement are time series, and we therefore use a Long Short-Term Memory (LSTM) neural network, which is a type of Recurrent Neural Network (RNN)~\cite{LSTM, Deep_learning}. RNNs consist of a unit cell that is repeated at every new input of the time-series data $\bold{x}^{(t)}$, producing an output $\bold{h}^{(t+1)}$ known as the hidden state. This hidden state is then combined with the next time-series input $\bold{x}^{(t+1)}$, allowing information to propagate through the sequence and have an impact on the outputs at future times (i.e., acting as a memory)~\cite{Deep_learning}. LSTMs, in addition to a hidden state, use a cell state $\bold{c}^{(t)}$ to retain values for arbitrarily long periods of time~\cite{LSTM}. Indeed, the units of a LSTM are composed of three gates (see Fig.~\ref{model_illu}): an input gate, an output gate, and a forget gate, to determine which information from the prior hidden state must be taken into account, stored and erased respectively. Therefore, this architecture is specifically designed to deal with long-time dependencies in sequential data. The architecture of the model we implemented is illustrated in Fig.~\ref{model_illu}.
The input layer of our network is a LSTM one with $100$ units in its hidden state, to take as input the time series data that are the photo currents. We use the last hidden state along the sequence length (dimension $100$) as the input of a $40$-neuron dense layer with activation function set to ReLU. The output layer is a $4$-neuron linear layer. The loss function we use for training is the sparse categorical cross entropy since we deal with a $4$-class classification problem, and the optimizer is Adam with default parameters. The model is trained on $9\times10^{4}$ photo currents, with dropout to reduce overfitting, and tested on $10^{4}$ photo currents. Before being fed into the network, the photocurrent values at each time step are standardized to have zero mean and unit variance.\\

Hence, our model takes as input the homodyne photo currents the spy obtains while monitoring the photons, and produces a 4-dimensional score vector (i.e., logits) over the four possible initial states of the BB84 protocol (i.e., $\ket{0}, \ket{1}, \ket{+}, \ket{-}$), which can be interpreted as a probability distribution after applying a softmax transformation. This supposes that the spy has the ability to train the neural network beforehand, using a similar photon source, optical fiber and detector than Alice and Bob, which is not unrealistic since one could assume the spy know which kind of QKD devices Alice and Bob bought on the available market.

\subsection{Impact of measurement VS Accuracy}

As explained earlier, the goal of the spy is to minimize their impact on the photon states while maximizing the part of the sifted key obtained.  To quantify the impact of the measurement, we use the QBER introduced earlier [Eq.~(\ref{eqQBER})], as it is directly measurable by Alice and Bob and allows them to assess the security of the protocol. To quantify the success of the spy in retrieving the initial state of the photons, we use the spy accuracy that we will denote by $A$. In the context of our deep learning approach, this is the neural network test accuracy~\cite{Deep_learning} which is defined as the percentage of good predictions among all the predictions of the network on the test set. Although there are several metrics used in machine learning (e.g., F1-score), accuracy reflects the model percentage of success in a given task, and suits our problem given the four initial states are equiprobable.\\
When dealing with a projective measurement, this accuracy is defined as the probability that the spy measures the right state.
Note that the amount of information extracted from the qubits (e.g., the information gain) is defined by the measurement scheme. On the other hand, accuracy $A$ depends on both the information gain and the ability of the neural network to efficiently harness it, as detailed in Sec.~\ref{information_gain}. Thus $A$ does not correspond properly to a measure of the extracted information in the sense of \textit{Shannon}~\cite{Shannon_1948}. 
However, we use it for convenience to quantify the success of the eavesdropping scheme, since it is a \textit{performance} measure that represents the average percentage of the sifted key obtained by the spy.
The accuracy as defined is a performance measure for the $4$-states classification task, and does not entirely reflect the real performance of the spy, which is to obtain the sifted key, i.e., a $2$-class classification problem. For this purpose, the key accuracy $A_{key}$ is employed here.

\section{Results}\label{results}

In this section, we study the impact of our attack and its performance in different cases in terms of QBER and accuracy $A$. We first compute the QBER in the case of no attack. Then, we study a simple standard projective measurement attack, before investigating our continuous measurement scheme. Note that in this section, except stated otherwise, all time durations are measured in units of $1/\omega$.
\subsection{No attack} 

In the case where no spying is done on the quantum channel, which means there is no measurement and only the intrinsic dissipation, the QBER can easily be obtained from Eq.~(\ref{SME}) with $e = 0$, which corresponds to a Lindblad master equation (see Appendix~\ref{A1}), and reads
\begin{equation}\label{QBERe0}
    \mathrm{QBER} = \frac{1}{4} - \frac{e^{-2\gamma_D t_f}}{4}, \quad \quad 0 \leqslant \mathrm{QBER} \leqslant 25\%,
\end{equation}
where $t_f$ is the total travel time of the qubit in the noisy quantum channel. Hence, the QBER ranges from $0$ for a perfect channel to $25\%$ for a very noisy channel or a very long travel time

\subsection{Attack via projective measurement}

Let us now consider that the spy performs a projective measurement on the qubit at a certain time $t^*$ ($0 < t^* < t_f$), as in an Intercept-and-Resend attack. Like Bob, the spy does not know in advance which measurement basis he should use, and thus measures randomly in the Pauli-X or Pauli-Z bases.

In this case, the accuracy $A$ can be calculated exactly by solving the Lindblad master equation with a single jump operator $L=\sigma_x$ (see Appendix~\ref{B}) and reads 
\begin{equation}
    A = \frac{5}{8} + \frac{e^{-2\gamma_Dt^*}}{8}, \quad\quad 62.5\% \leqslant A \leqslant 75\%
    \label{proba_eve}
\end{equation}
which depends on the time $t^*$ at which the projective measurement is performed. Therefore, Eve must measure the photons as close to Alice as possible in order to maximize Eq.~(\ref{proba_eve}) and get as much as possible of the sifted key, which is here bounded by $75\%$, meaning that the spy has at best $75\%$ chance to guess the initial state sent by Alice.\\

The QBER can also be obtained easily (see Appendix~\ref{A2}), and reads
\begin{equation}\label{QBEReproj}
    \mathrm{QBER} = \frac{3}{8} - \frac{e^{-2\gamma_D t_f}}{8}, \quad \quad 25\% \leqslant \mathrm{QBER} \leqslant 37.5\%,
\end{equation}
Interestingly, we see that the time $t^*$ at which Eve performs her measurement does not impact the probability that Bob measures the state he is supposed to. Also, comparing Eqs.~(\ref{QBERe0}) and (\ref{QBEReproj}), we clearly see that Alice and Bob will easily distinguish the presence of the spy from intrinsic dissipation.

\subsection{Attack via continuous measurement}

We now discuss our new kind of attack, based on an homodyne measurement of the photon that is fed to a LSTM neural network. When modeling the dynamics of photons under homodyne detection, one must set the measurement operator $e$ of Eq.~(\ref{SME}). First, let us use \begin{equation}
\begin{aligned}
    e &= \sigma_z,\\
\end{aligned}
\end{equation}
and consider in the first instance that the homodyne measurement is performed during the whole travel time, set to $\omega t_f = 0.1$, and with other parameters $\eta = 1$, $\gamma_E = 5\gamma_D = 5\omega$.
With these parameters we obtain, from the solutions of Eq.~(\ref{SME}), a QBER of $20.5\%$, lower than the $37\%$ of the projective measurement obtained from Eq.~(\ref{QBEReproj}) and higher than the $4.5\%$ of the case with dissipation only, obtained from Eq.~(\ref{QBERe0}). In addition, we get a neural network test accuracy $A \approx 43\%$, which is below the interval given by Eq.~(\ref{proba_eve}). Hence, we see that the spying accuracy achieved via this simple continuous measurement scheme is lower than the one achieved via the projective measurement, but the QBER is lower.\\

In an attempt to reduce the impact of the spy while increasing its effectiveness, we now parameterize the measurement operator $e$ as depending on an angle $\theta$ as
\begin{equation}
    e = \cos(\theta)\sigma_x + \sin(\theta)\sigma_z,
    \label{theta_ope}
\end{equation}
so that it corresponds to a superposition of the two polarization bases. 

\begin{figure}
    \centering
    \includegraphics[width = 0.975\linewidth]{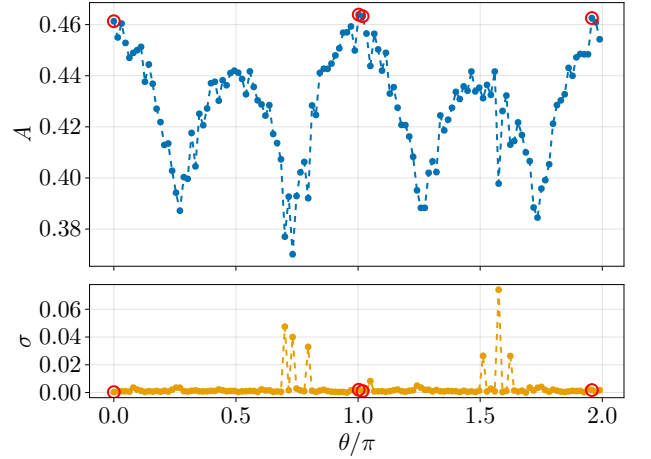}
    \caption{\justifying{Mean estimated accuracy $A$ (blue) and standard deviation (orange) of the model on the test set as a function of $\theta$. The photo currents of the test set were obtained using Eq.~(\ref{SME}) and~(\ref{Jhom}). Circled in red are the four maximum accuracy values and their corresponding standard deviations, reported in Table I. Other parameters are $\omega t_f = 0.1$ and $\gamma_E = 5\omega = 5\gamma_D$.}}
    \label{accuracy_theta}
\end{figure}

Let us first look at the accuracy $A$ yielded by this new measurement operator as a function of $\theta/\pi$, which is depicted in Fig.~\ref{accuracy_theta}, together with the associated standard deviation. There are four angles leading to an accuracy around $46\%$, as summarized in Table~\ref{tab_acc_theta}, which is higher than the $43\%$ found earlier. Note that the four angles seem equivalent given their values and the standard deviations.

\begin{table}
\centering
\begin{tabular}{ccc}
    \toprule
     $\theta$ & mean accuracy & standard deviation \\
    \midrule
    $0$ & $46.1\%$ & $0.03\%$      \\
    $\pi$ & $46.4\%$ & $0.2\%$     \\
    1.02$\pi$ & $46.3\%$ & $0.1\%$ \\
    1.96$\pi$ & $46.2\%$ & $0.2\%$ \\
    \bottomrule
\end{tabular}
\caption{\justifying{Mean accuracy $A$ of the neural network and corresponding standard deviations for the optimal values of $\theta$ found in Fig.~\ref{accuracy_theta}.}}
\label{tab_acc_theta}
\end{table}

\begin{figure}
    \centering
    \includegraphics[width = 0.95\linewidth]{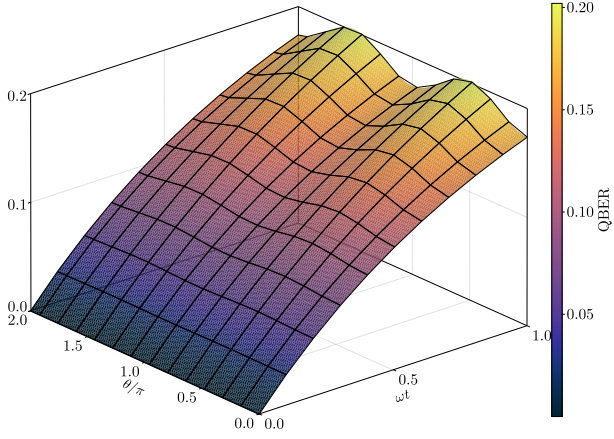}
    \caption{\justifying{QBER as a function of time and measurement angle $\theta$. The evolution of the photon states through time was obtained using Eq.~(\ref{SME}). Other parameters are $\omega t_f = 0.1$ and $\gamma_E = 5\omega = 5\gamma_D$.}}
    \label{avg_qber}
\end{figure}

\begin{figure}
    \centering
    \includegraphics[width = 0.475\textwidth]{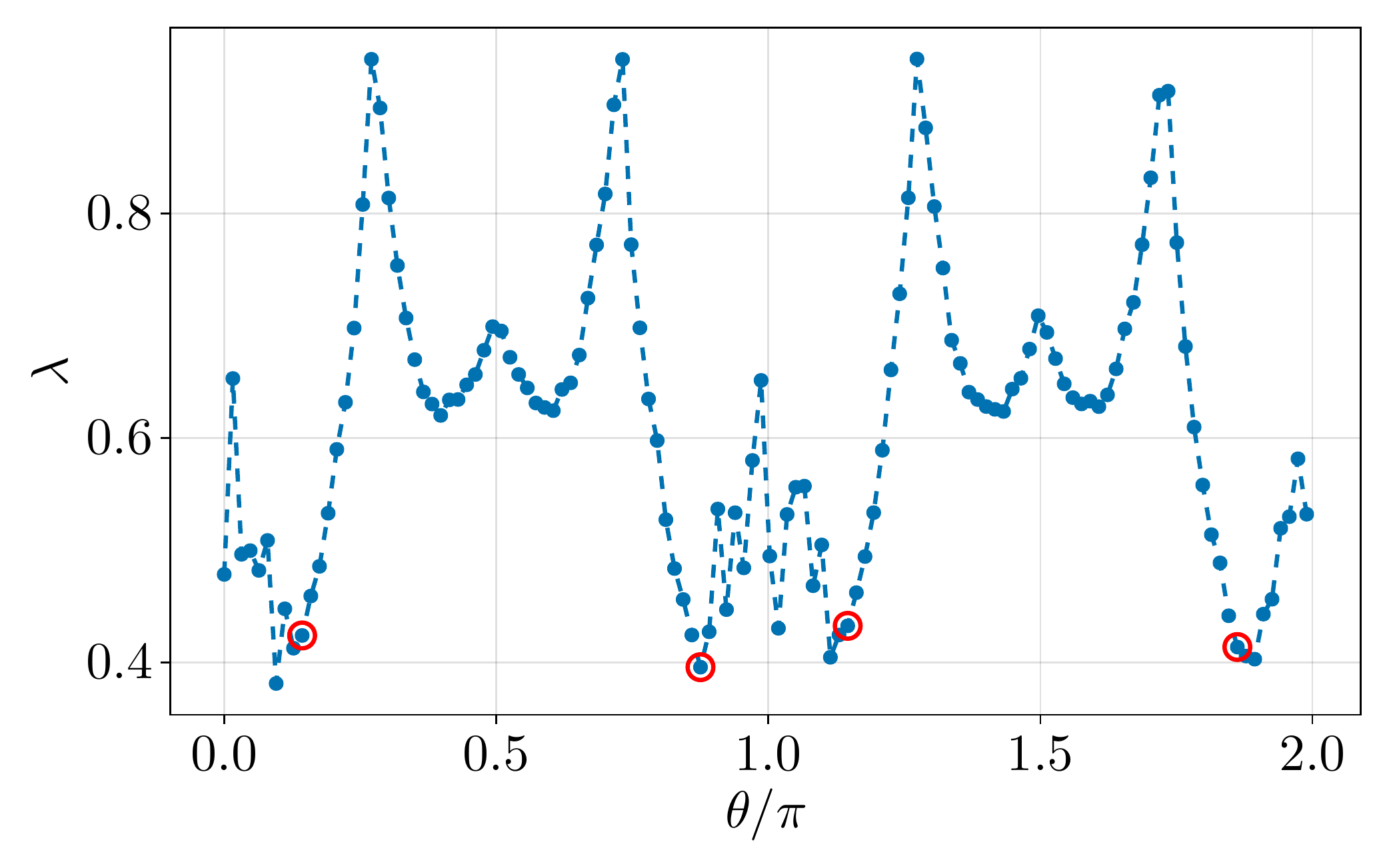}
    \caption{\justifying{$\lambda(\theta)$ [Eq.~(\ref{lambda_eq})] as a function of $\theta$. Circled in red are the four values of $\theta$ maximizing $A$ found in Fig.~\ref{accuracy_theta} and reported in Table I. Other parameters are $\omega t_f = 0.1$ and $\gamma_E = 5\omega = 5\gamma_D$.}}
    \label{lambda}
\end{figure}

To obtain the impact of the measurement on the BB84 protocol itself, we average the QBER over the four possible initial states, and evaluate it as a function of $\theta$ and $\omega t$, as displayed in Fig.~\ref{avg_qber}.
One can see there is a trade-off between the accuracy and the disturbance that the measurement induces. However, the angles that minimize the QBER, which are $\theta = 0 \pm k\pi , k\in \mathbb{Z}$, yield a better test accuracy, thus increasing the spy accuracy (see Fig.~\ref{accuracy_theta}). In order to quantify this tradeoff, we define a new quantity $\lambda(\theta)$ as the QBER divided by the accuracy $A$ of the network for a given measurement basis (i.e.,  a given $\theta$)
\begin{equation}
    \lambda(\theta) = \frac{\mathrm{QBER}(\theta)}{A(\theta)},
    \label{lambda_eq}
\end{equation}
which is shown for $\omega t_f = 0.1$ in Fig.~\ref{lambda}. As expected, we observe that among the four measurement angles maximizing the spy accuracy (circled in red), $\theta =\pi$ yields the lowest $\lambda$ ratio, with an accuracy around $46.5\%$ and a QBER around $17.5\%$, though the other angles give similar performances.

\begin{figure}
    \centering
    \includegraphics[width = 0.975\linewidth]{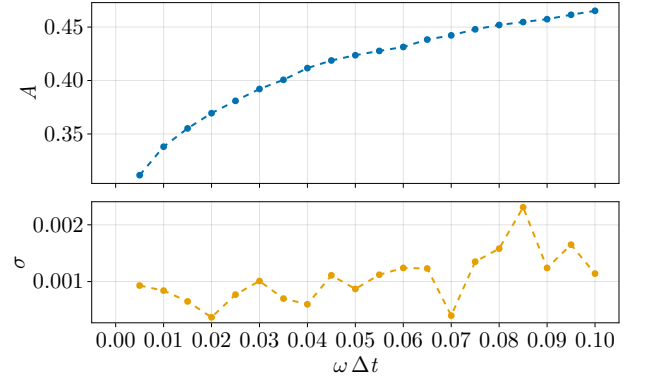}
    \caption{\justifying{Mean accuracy $A$ and standard deviation of the neural network as a function of the measurement length $\omega \Delta t$. The photo currents were obtained using Eq.~(\ref{SME}) and~(\ref{Jhom}). Other parameters are $\omega t_f = 0.1$ and $\gamma_E = 5\omega = 5\gamma_D$.}}
    \label{acc_vs_length}
\end{figure}

Finally, we analyze the impact of the measurement duration (denoted $\omega \Delta t$) on the performance of the attack. Indeed, one could expect the information about the initial state to be mostly contained in the early stages of the currents, thus allowing to decrease its duration and its impact on the qubit states while maintaining a reasonable accuracy. We choose here the optimal measurement angle found earlier, $\theta = \pi$. As shown in Fig.~\ref{acc_vs_length}, which displays $A$ as a function of $\omega \Delta t$, the accuracy reaches $44\%$ by only measuring until $\omega\Delta t = 0.07$ while the QBER decreases to $14.5\%$.

Altogether, taking $\theta = \pi$, the optimal duration $\omega \Delta t = 0.07$, and increasing $\gamma_E$ to $10\omega$ we obtain
\begin{align}\label{bestA}
&A = 47.5\% ,\\\label{BestQBER}
&\mathrm{QBER} = 20\%,
\end{align}
the latter representing a $15\%$ increase compared to the time-evolved state where no measurement is made, as summarized in Table~\ref{tab_means_std}.
However, this accuracy over the four initial states represents an accuracy over the key bits $A_{key} = 73\%$.

\begin{table*}[ht!]
        \begin{center}
            \begin{tabular}{c|c|c|c}
            \toprule
            Dissipation & Attack & Impact (QBER) & Key accuracy \\
            \midrule
            $d = 0$ & $e = 0$ & $0.0$     & $0.0$  \\
            $d = \sigma_x$ & $e=0$ & $4.5\%$    & $0.0$   \\
            \midrule
             $d = \sigma_x$ & Projective measurement & $27.2\%$ &  $74.1\%$\\
            \midrule
             & DLCA &&\\
             $d = \sigma_x$ & $e=\sigma_z\ (\gamma_E = 5\omega)$ & $20.5\%$ & $68\%$  \\
             $d = \sigma_x$& $e = -\sigma_x\ (\gamma_E = 10\omega)$&  $20\%$ &  $73\%$\\
             &($\theta = \pi$)&&\\
            \bottomrule
            \end{tabular}
        \caption{\justifying{Summary of the QBER and accuracies $A_{key}$ generated by the different attack schemes we analyzed. The parameters are $\omega t_f = 0.1$, $\omega \Delta t = 0.07$, and $\omega = \gamma_D$. We set Eve measurement time $\omega t^*$ to $0.35$ such that it corresponds to the middle of the optimized continuous measurement.}}
        \label{tab_means_std}
        \end{center}
    \end{table*}

\section{Information gain and key rates}\label{info_gain_key_rate}

So far, we have presented the performances of our attack in terms of QBER and accuracy $A$. We now discuss how our attack fits into the existing security proofs and how it compares to optimal individual attacks in terms of information gain. Finally, we close the section by evaluating the typical key rates our attack yields.

\subsection{Information-Disturbance Principle}\label{information_gain}
As predicted by the laws of quantum mechanics, it is impossible to gather information about the identity of a quantum system's state (when prepared in one of a set of non-orthogonal states) without introducing disturbance in said system~\cite{Fuchs_1996b}. From this, the information-disturbance principle establishes a trade-off between information gained from a measurement and the disturbance caused.
For the BB84 protocol, Shor and Preskill suggested in~\cite{2000_Shor_Preskill} that the maximum (Shannon) information Eve can have about the final key, per bit before privacy amplification, is
\begin{equation}
    I_{Eve} \leq H_2(e_x) + H_2(e_z),
    \label{info_bound}
\end{equation}
where $e_x$ ($e_z$) is the bit (phase) error rate, i.e., the rate of errors in the Z-basis (X-basis), and 
\begin{equation}
    H_2(p) = -p\log_2(p)-(1-p)\log_2(1-p)
\end{equation}
is the binary entropy function.

An estimator of $I_{Eve}$ is the information gain, or equivalently the expected mutual information, which is defined for two continuous random variables $X$ and $Y$ as 
\begin{equation}
I(X; Y) = \iint P_{X,Y}(x, y) \log \left( \frac{P_{X,Y}(x,y)}{P_X(x) P_Y(y)} \right) dx\,dy.
\end{equation}
Let $S$ be a discrete variable representing the different initial states ($s=0,1,2,3$) and $X\in\mathbb{R}^{70}$ a $70$-dimensional continuous variable representing the values of the homodyne currents at each time step, we obtain (see Appendix~\ref{info_gain_dvlp}) that the $95\%$ confidence interval on the information gain yielded by the homodyne measurement, with the parameters of Eq.~(\ref{bestA}) and~(\ref{BestQBER}), on the qubits of the BB84 protocol is
\begin{align}
    I(S;x) = \left[0.1519,\ 0.1823\right]\ \text{bits},
    \label{info_gain}
\end{align}
which represents information about the initial qubit states in $\left[0.3880,\ 0.4275 \right]$ bits. Also, one can compute the mutual information between Eve's and Alice's keys from the confusion matrix of the deep learning model (see Appendix~\ref{info_gain_dvlp}), and obtain $I(A;E) = 0.1527$ bits. Thus with $95\%$ confidence, the model saturates the data processing inequality, proving the only limitation of the DLCA attack arises from the homodyne measurement itself and the information it extracts.

At the same time, the estimated bit and phase error rates being, from Eq.~(\ref{BestQBER}), $e_x = 0.20$ and $e_z = 0.20$, Eq.~(\ref{info_bound}) becomes 
\begin{equation}
    I_{Eve} \leq 1.44\ \mathrm{bits},
    \label{info_bound_final}
\end{equation}
showing that Eq.~(\ref{info_gain})  is below the known threshold. Note that this latter is above the $1$ bit of entropy in a single key bit for our specific choice of parameters, which means that the final key rate of the protocol would be negative, such that no secure key can be distilled under standard BB84 security assumptions. This is a consequence of the QBER being above the $11\%$ threshold obtained by Shor and Preskill~\cite{2000_Shor_Preskill}, i.e., the threshold above which the security of the protocol cannot be guaranteed.

If we now consider a regime below the $11\%$ threshold and analyze the information gained by the homodyne measurement when no dissipation is occurring in the optical fiber (such that all of the QBER is caused by the spy), we obtain (with $\gamma_D = 0$ and $\gamma_E = 4\omega$)
\begin{align}
    &\mathrm{QBER} = 10.7\% \leq 11\%, \\
    &A_{key} = 69\% ,  \\
    &I(S;x) \in \left[0.0919,\ 0.1222\right]\ \mathrm{bits} \approx I_{Eve} \label{I_no_noise} \\
    &I_{Eve} \leq H_2(0.107) + H_2(0.107) = 0.98 \ \mathrm{bits}.
\end{align}
As $I(S;x) \leq 0.98\ \mathrm{bits}$, this shows that the DLCA still respects the established theoretical bounds.

\paragraph{Comparison with an optimal individual attack.} So far, we have compared our attack to the unconditional bound [Eq.~(\ref{info_bound})]. Here, we restrict Eve's power to unitary individual attacks only, for which the information-disturbance
principle reads~\cite{Fuchs_1997}
\begin{equation}\label{boundI}
    I(A;E)\leq \frac{1}{2}\phi\left[2\sqrt{\mathrm{QBER}(1-\mathrm{QBER})}\right],
\end{equation}
where $I(A;E)$ is the mutual information between Alice and Eve and $\phi[z]\equiv (1+z)\ln(1+z) + (1-z)\ln(1-z)$. It has been shown that such a bound can be saturated using a phase-covariant quantum cloner, which is an approximate cloning procedure for two-level systems on the equator of the Bloch sphere~\cite{Bruss_2000}, and that a secure key can be distilled as long as the QBER is below $14.65\%$, point at which the curves $I(A;E)$ and $I(A;B)$ intersect. Inserting this value into Eq.~(\ref{boundI}), we obtain that Eve's maximum obtainable information is $0.399$ bits. 
With such QBER, the DLCA attack reaches $A = 45.5\%,\ A_{key} = 70.7\%$, and the homodyne measurement estimated information gain lies in $\left[0.1204, 0.1511\right]$ bits with $95\%$ confidence. Also, the mutual information between Eve's and Alice's key is $0.127$ bits.

\subsection{Key rates} \label{sec_key_rate}

The key rate of a QKD protocol is defined as the percentage of secure key bits which can be extracted from the sifted key. It basically makes it possible to calculate the amount of bits Alice and Bob must sacrifice in the error correction and privacy amplification procedure in order to obtain a secure key.
Devetak and Winter, in~\cite{Devetak_2005}, demonstrated the following general and composable bound on the key rate of QKD protocols
\begin{equation}
    R \geq I(A;B) - I(A;E),
    \label{devetak_rate}
\end{equation}
where the first term quantifies the error correction cost while the second one quantifies how much privacy amplification is needed.
Since the mutual information between two random variables $X$ and $Y$ can be expressed as
\begin{equation}
    I(X;Y) = H_2(X) - H_2(X\mid Y) = H_2(Y) - H_2(Y\mid X),
    \label{mutual_info}
\end{equation} 
we obtain 
\begin{equation}
    R \geq H_2(A\mid E) - H_2(A\mid B),
\end{equation}
which is saturated in the asymptotic limit on infinitely long keys~\cite{Diamanti_2015}.
Below, we evaluate the typical key rates our attack yields under a more realistic noise model: the depolarizing channel model. 

\paragraph{Depolarizing channel model.}
So far in this paper we have considered a toy model for the intrinsic dissipation occurring in the optical fiber, in order notably to understand the effect of anisotropic dissipation. However, the results presented here could be straightforwardly generalized to more realistic and complex noise models. One such model is the depolarizing quantum channel, an isotropic noise model often used as a simple and effective way to represent noise in quantum communication~\cite{Nielsen_Chuang}. For a single qubit, the depolarizing channel is, in a Lindblad form,
\begin{equation}\mathcal{D}_{\mathrm{depol}}(\rho)=\frac{\gamma_D}{3}\left(\mathcal{D}\left[\sigma_x\right](\rho)+\mathcal{D}\left[\sigma_y\right](\rho)+\mathcal{D}\left[\sigma_z\right](\rho)\right).
\label{depolarization}
\end{equation}
\paragraph{DLCA attack.}
By using the dissipator~(\ref{depolarization}) with $\gamma_D =\omega/100$, $\gamma_E = 7\omega/5$ and taking the optimal angle and measurement duration found above, Eve's neural network accuracy about Alice's key becomes $68\%$ and the QBER $10.9\%$. Thus, the conditional entropies, per bit, of Alice's bit given Eve's and Bob's are
\begin{align}
    H_2(A\mid E) &= H_2(0.68) = 0.90\ \mathrm{bits},\\
    H_2(A\mid B) &= H_2(1-0.11) = 0.497\ \mathrm{bits},
\end{align}
which yields a final key rate of
\begin{equation}
    R = 0.403\ \mathrm{bits}.
    \label{key_rate}
\end{equation}
Eq.~(\ref{key_rate}) constitutes an upper bound on the usable key rate for Alice and Bob. Indeed, if Alice and Bob make the most pessimistic assumption that the whole QBER of $10.9\%$ is generated by eavesdropping, the key rate they obtain is $0.006$ bits.

\section{Conclusion}\label{conclu}
In this paper, we introduced a new type of individual attack on QKD protocols based on continuous measurement that, used as an input of a trained recurrent neural network, allows the spy to retrieve with high accuracy the sifted key bits sent by one of the parties without being significantly noticed. We denote our attack as a \textit{Deep-learning-based continuous attack} (DLCA). Although more quantitative and comparison analyses should be done, also in terms of noise models considered for the optical fiber, our attack scheme exhibits better performances than a projective measurement attack. Note that since we assume the use of perfect single photon sources and detectors, our attack does not compromise the security of the BB84 protocol as long as Alice and Bob perform enough privacy amplification, reducing the key rate at least below the upper bound we computed. However, our attack could be adapted to quantum hacking setups targeting vulnerabilities in the implementation of QKD protocols likely to compromise their security. Our work constitutes a first step towards this goal, as well as other promising research directions outlined below.  

Indeed, to go further, one could for example investigate the possible generalization of our strategy to a collective or coherent attack~\cite{Gisin_2002}.

Also, a more complex and realistic noise model of the optical fiber could be used. In~\cite{Kozubov2019}, Kozubov \textit{et al.} span the space using three states: the vacuum state and the states we denoted $\ket{0}$ and $\ket{1}$ in this work (i.e., horizontally and vertically polarized photons). By doing so, they take into account the non-zero probability that the photon is absorbed in the optical fiber. They also tune the phenomenological parameters involved in the master equation, which allows to take into account the phenomena of birefringence, isotropic absorption, and dichroism. One could also consider the potential losses caused by the imperfection of Alice and Bob detectors. Overall, it is straightforward to adapt our approach to such other noise models.

In addition, one could investigate practical implementations of our attack scheme involving homodyne detection of single photon~\cite{Lvovsky_2009}, by exploiting evanescent waves in optical fibers~\cite{Bertolotti_2017} or quantum memories~\cite{Lvovsky2009}, as we partially started in Sec.~\ref{adia}.

One could also investigate how our scheme could be applied to decoy states protocols~\cite{Decoy_state,Lo_2005}, coherent states continuous variable protocols~\cite{Zhang_2024, Jain2022} exploiting homodyne or heterodyne detection, or entanglement-based protocols such as the E91~\cite{Ekert_1991}, the BBM92~\cite{Bennett_1992} or on device-independent protocols~\cite{Antonio_2007,Zhang_2022}. Entanglement-based protocols are promising for satellites QKD, which is currently being extensively studied by the scientific community~\cite{Liao_2017, Ecker_2022, Yin_2017, Chen_2021}. One could thus explore the generalization of our attack and its practical implementation to satellite QKD.

Also, one could consider that Eve uses quantum feedback based on the measurement outcomes to try to cover her tracks. Depending on the noise model and the regime of parameters, such a conditional feedback could yield Non-Markovian dynamics which could be studied via a Non-Markovian approach such as cHEOM~\cite{Link_2022}.

Finally, one could also analyze our attack in the context of QKD protocols using qudits, also called high dimensional quantum key distributions (HDQKD)~\cite{Zahidy_2024, Yan_2019, halevi_2024}.

\begin{acknowledgments}
We thank Jérôme Denis and John Martin for helpful initial discussions on the topic. More particularly, we express our sincere thanks to Jérôme Denis for asking the question that inspired this work and for his valuable assistance in the elaboration of the figures.
We also thank Daniel Oi for his insightful inputs on information-theoretics-QKD.
\end{acknowledgments}
\appendix

\section{Homodyne detection schemes} \label{adia}

In this section, we derive the stochastic master equation~(\ref{SME}) starting from the one modeling the homodyne detection of a damped ancillary field that couples to the photons in the optical fiber, via e.g.\ their evanescent waves~\cite{Bertolotti_2017}.

We consider the stochastic master equation for the full density operator $\rho$ of the combined system made of an optical fiber photon and an ancillary field of the form
\begin{equation}
\begin{aligned}
        d\rho= &-i\left[H + \omega_a a^\dagger a + i g(e a^\dagger - a e^\dagger),\rho\right]dt + \gamma_D\mathcal{D}[d]\rho dt \\
        &+ \gamma_a\mathcal{D}[a]\rho dt + \sqrt{\gamma_a\eta}\mathcal{H}[a]\rho dW,
        \label{SME0}
\end{aligned}
\end{equation}
where $a$ ($a^\dagger$) is the annihilation (creation) operator for the ancillary field of frequency $\omega_a$ damped with a rate $\gamma_a$. The Heisenberg equation of motion for $a$ reads
\begin{equation}
    \dot{a} = -(i\omega_a + \gamma_a) a + g e.
\end{equation}
For $\gamma_a \gg g, \omega_a, \gamma_D$, the ancillary field remains weakly populated and can be adiabatically eliminated, as in the bad cavity limit in cavity/circuit QED. According to this, the state of the ancillary field relaxes rapidly and we can set the left-hand-side of the equation above to zero. This makes it possible to slave the ancillary field to the photonic degrees of freedom:
\begin{equation}\label{adiaa}
    a \approx \frac{g}{\gamma_a} e.
\end{equation}
Replacing $a$ in Eq.~(\ref{SME0}) by Eq.~(\ref{adiaa}) directly yields the stochastic master equation~(\ref{SME}) of the main text with $\gamma_E = g^2/\gamma_a$.

An alternative implementation of the attack would consist in tapping a small fraction of the quantum signal using a low-reflectivity beam splitter, or, in a photonic integrated circuit (PIC) scenario, a directional coupler. The weakly extracted component would then be interfered with a strong local oscillator via a second beam splitter (or coupler), enabling standard homodyne detection of a chosen quadrature.

We acknowledge the practical challenges associated with realizing this attack using current technology. Among them, the most constraining is arguably the requirement for high-bandwidth, low-noise detection—potentially in the GHz range—to resolve the short temporal modes used in state-of-the-art QKD systems. Nonetheless, the performance of photodetectors and associated readout electronics has improved significantly over the past decades, particularly in terms of bandwidth, quantum efficiency, and noise suppression. Crucially, there are no known fundamental physical limits that prevent further improvements in these areas. We thus believe the attack strategies proposed here are not only conceptually valid, but also increasingly realistic in light of technological trends.

\section{Analytical derivation of the QBER without attacks} \label{A1}

Without measurement, we model the evolution of a photon state in the optical fiber with the Lindblad master equation
\begin{equation}
    \Dot{\rho} = -i\left[H,\rho\right] + \gamma_D\left(L\rho L^\dag - \frac{1}{2} L^\dag L\rho - \frac{1}{2}\rho L^\dag L\right),
    \label{Lindblad_final}
\end{equation}
where the jump operator $L=\sigma_x$ models bit flip errors, and the Hamiltonian is $H=\omega\sigma_z$ for the initial states $\ket{0}$ and $\ket{1}$ and $H=\omega\sigma_x$ for the initial states $\ket{+}$ and $\ket{-}$. 
We denote the matrix elements of $\rho$ in the basis $\{ \ket{0}, \ket{1}\}$ as
$\rho_{ij} = Tr \left(\ket{j} \bra{i} \rho \right) = \bra{i} \rho \ket{j} (i,j = 0,1)$. Projecting the master equation in the computational basis gives the following linear set of equations for the density matrix elements
 \begin{equation}
    \begin{cases}
        \Dot{\rho}_{00} = \gamma_D(\rho_{11}(t) - \rho_{00}(t)) \\
        \Dot{\rho}_{01} = \gamma_D(\rho_{10}(t) - \rho_{01}(t)) \\
        \Dot{\rho}_{10} = \gamma_D(\rho_{01}(t) - \rho_{10}(t)) \\
        \Dot{\rho}_{11} = \gamma_D(\rho_{00}(t) - \rho_{11}(t)) \\
    \end{cases}  
    \end{equation}
    which is independent of the Hamiltonian term of the master equation for both cases $H =\omega \sigma_x$ and $H =\omega \sigma_z$.
    Resolving this system gives: \begin{widetext}
    \begin{equation}
        \rho(t) = 
    \begin{pmatrix}
        \frac{e^{-2\gamma_D t}}{2}(1+e^{2\gamma_D t})\rho_{00}(0) + \frac{e^{-2\gamma_D t}}{2}(-1+e^{2\gamma_D t})\rho_{11}(0) & \frac{e^{-2\gamma_D t}}{2}(1+e^{2\gamma_D t})\rho_{01}(0) + \frac{e^{-2\gamma_D t}}{2}(-1+e^{2\gamma_D t})\rho_{10}(0)\\
        &\\
        \frac{e^{-2\gamma_D t}}{2}(-1+e^{2\gamma_D t})\rho_{01}(0) + \frac{e^{-2\gamma_D t}}{2}(1+e^{2\gamma_D t})\rho_{10}(0) & \frac{e^{-2\gamma_D t}}{2}(-1+e^{2\gamma_D t})\rho_{00}(0) + \frac{e^{-2\gamma_D t}}{2}(1+e^{2\gamma_D t})\rho_{11}(0)\\
    \end{pmatrix}
    \label{evolution_equation}
    \end{equation}
\end{widetext}
which describes the state of the qubit in the channel at time $t$.

There are four possible states for Alice to send~:$\{\ket{0},\ket{1},\ket{+},\ket{-}\}$.\\
In the case $\rho(0) = \ket{0}\bra{0}$, Eq.~(\ref{evolution_equation}) gives  
\begin{equation}
    \rho(t) = \begin{pmatrix}
    \frac{e^{-2\gamma_D t}}{2} + \frac{1}{2} & 0\\
    0 &  \frac{-e^{-2\gamma_D t}}{2} + \frac{1}{2}\\
\end{pmatrix}
\end{equation}
so that the probability that the qubit is in the state $\ket{0}$ after going through the optical fiber is $\rho_{00}(t) = \frac{e^{-2\gamma_D t}}{2} + \frac{1}{2}$.

In the case $\rho(0) = \ket{1}\bra{1}$, we find 
\begin{equation}
    \rho(t) = \begin{pmatrix}
    \frac{-e^{-2\gamma_D t}}{2} + \frac{1}{2} & 0\\
    0 &  \frac{e^{-2\gamma_D t}}{2} + \frac{1}{2}\\
\end{pmatrix}
\end{equation}
and the probability that the qubit is in the state $\ket{1}$ is given by $\rho_{11}(t) = \frac{e^{-2\gamma_D t}}{2} + \frac{1}{2}$.

In the case $\rho(0) = \ket{+}\bra{+}$, we find 
\begin{equation}
    \rho(t) = \begin{pmatrix}
    \frac{1}{2} &  \frac{1}{2}\\
    \frac{1}{2} &  \frac{1}{2}\\
\end{pmatrix}.
\end{equation}

Finally, in the case $\rho(0) = \ket{-}\bra{-}$, we find 
\begin{equation}
    \rho(t) = \begin{pmatrix}
    \frac{1}{2} &  \frac{-1}{2}\\
    \frac{-1}{2} &  \frac{1}{2}\\
\end{pmatrix}.
\end{equation}
As the states $\ket{+}$ and $\ket{-}$ are eigenstates of the $\sigma_x$ jump operator, they will not change when traveling through the optical fiber: they are insensitive to the dissipation process. These states are decoherence-free states (or dark states) for the given master equation.\\

When the photon reaches Bob at time $t_f$, he chooses randomly one of the two available bases (i.e., Pauli-X and Pauli-Z). The probabilities above correspond to Bob measuring in the right basis. On the other hand, the probability that Alice sends one of the four states is $\frac{1}{4}$ because she chooses randomly the polarization basis and the state in this polarization. Thus, the probability $\probP_b(\text{same results})$ that Bob gets the right state is
\begin{equation}
\begin{aligned}
    \probP_b(\text{same results})&= \frac{1}{4}\frac{1}{2}\left(\frac{e^{-2\gamma_D t_f}}{2} + \frac{1}{2}\right) + \frac{1}{4}\frac{1}{2}\left(\frac{e^{-2\gamma_D t_f}}{2} + \frac{1}{2}\right) \\
    &+ \frac{1}{4}\frac{1}{2}1 + \frac{1}{4}\frac{1}{2}1 \\
    &= \frac{e^{-2\gamma_D t_f}}{4} + \frac{3}{4}.
\end{aligned}
\label{accuracy_bob}
\end{equation}
Since it corresponds to one minus the QBER, we finally have
\begin{equation}
    \mathrm{QBER} = \frac{1}{4}-\frac{e^{-2\gamma_D t_f}}{4},
\end{equation}
which corresponds to Eq.~(\ref{QBERe0}) in the main text.

\section{Analytical derivation of Eve accuracy for a projective measurement} \label{B}

The result above allows us to easily determine Eve accuracy in the case she performs a projective measurement at time $t^*$. Indeed, when the photon reaches Eve at time $t^*$, she also chooses randomly one of the two available bases. The probabilities above correspond to someone measuring in the right basis. However, if Eve does not, her probability of detecting the right state is still 1/2 as she can get each result with equal probability. Thus, the probability that Eve deduces the right state (i.e., the accuracy $A$) can be obtained from Eq.~(\ref{accuracy_bob}), which yields
\begin{equation}
\begin{aligned}
    A &= \frac{1}{4}\frac{1}{2}\left(\frac{e^{-2\gamma_D t^*}}{2} + \frac{1}{2} + \frac{1}{2}\right) + \frac{1}{4}\frac{1}{2}\left(\frac{e^{-2\gamma_D t^*}}{2} + \frac{1}{2} + \frac{1}{2}\right) \\
    &\quad+ \frac{1}{4}\frac{1}{2}\left(1 + \frac{1}{2}\right) + \frac{1}{4}\frac{1}{2}\left(1 + \frac{1}{2}\right) \\
    &= \frac{e^{-2\gamma_D t^*}}{8} + \frac{5}{8},
    \label{eq25}
\end{aligned}
\end{equation}
which corresponds to Eq.~(\ref{proba_eve}) in the main text.

\section{Analytical derivation of Bob accuracy for a projective measurement (intercept-and-resend attack)} \label{A2}

After Eve's projective measurement at time $t^*$, the photon is in the state she measured, and thus evolves according to Eq.~(\ref{evolution_equation}) until it reaches Bob at time $t_f$. To simplify the process, we will look in detail to the case where Alice sends the initial state $\ket{0}$ which generalizes easily to the other states. Since we want to obtain the probability that Bob measures the state Alice sent, which is $1-\mathrm{QBER}$, and since they will both, at some part of the protocol, compare the bases they respectively used and discard the differing ones (see Sec.~\ref{BB84_steps}), we consider that Bob measures in the Pauli-Z basis $\{\ket{0},\ket{1}\}$.
There are four distinct cases, each corresponding to a different measurement result for Eve.\\
If she measures in the Pauli-Z basis (probability $1/2$) and she measures the state $\ket{0}$ [probability $(1 +e^{-2\gamma_Dt^*})/2$], the photon will be in the state $\ket{0}$ right after. Thus Bob will measure the state $\ket{0}$ with probability $(1 +e^{-2\gamma_D(t_f-t^*)})/2$.\\
However, if Eve measures in the Pauli-Z basis but the result is $\ket{1}$ [probability $(1-e^{-2\gamma_Dt^*})/2$] then the probability that Bob measures the state $\ket{0}$ is  $(1-e^{-2\gamma_D(t_f-t^*)})/2$.\\
If Eve measures in the Pauli-X basis (probability $1/2$), the result will be either $\ket{+}$ or $\ket{-}$, each with probability $1/2$, which is the state that will reach Bob since they are not affected by the dissipation. Therefore, Bob's result will be $\ket{0}$ or $\ket{1}$, each with probability $1/2$.\\
Altogether, this yields the probability that Bob measures the state $\ket{0}$ if Alice sent it
\begin{equation}
\begin{aligned}
        \probP_b(\ket{0}) &= \frac{1}{2}\left(\frac{1}{2}+\frac{e^{-2\gamma_Dt^*}}{2}\right)\left(\frac{1}{2}+\frac{e^{-2\gamma_D(t_f - t^*)}}{2}\right) \\
        &+ \frac{1}{2}\left(\frac{1}{2}-\frac{e^{-2\gamma_Dt^*}}{2}\right)\left(\frac{1}{2}-\frac{e^{-2\gamma_D(t_f - t*)}}{2}\right) \\
        &+ 2 \times \frac{1}{2} \times \frac{1}{2} \times\frac{1}{2}\\
        &= \frac{2}{4} + \frac{e^{-2\gamma_Dt_f}}{4}.
\end{aligned}
\end{equation}
Following the same procedure for the three other initial states, we obtain
\begin{align}
        \probP_b(\ket{1}) &= \frac{2}{4} + \frac{e^{-2\gamma_Dt_f}}{4}, \\
        \probP_b(\ket{+}) &= \probP_b(\ket{-}) = \frac{1}{2}\times 1 \times 1 + 2 \times\frac{1}{2} \times\frac{1}{2} \times\frac{1}{2} \nonumber\\
        & = \frac{3}{4}.
\end{align}
Since Alice sends each of these states with equal probability,
the final probability is 
\begin{equation}
    \begin{aligned}
        \probP_b (\text{right result}) &= \frac{1}{4}\probP_b(\ket{0}) + \frac{1}{4}\probP_b(\ket{1}) \\
        &+ \frac{1}{4}\probP_b(\ket{+}) + \frac{1}{4}\probP_b(\ket{-}) \\
        &= \frac{5}{8} + \frac{e^{-2\gamma_Dt_f}}{8}.
    \end{aligned}
\end{equation}
Since it corresponds to one minus the QBER, we finally have
\begin{equation}
    \mathrm{QBER} = \frac{3}{8} - \frac{e^{-2\gamma_D t_f}}{8},
\end{equation}
which corresponds to Eq.~(\ref{QBEReproj}) in the main text.

\section{Information gain computation}\label{info_gain_dvlp}
\paragraph{Information gain from the homodyne currents}
We start from the general definition of expected mutual information (i.e., information gain) for two random variables $X$ and $Y$~\cite{Shannon_1948, Wiley_Sons_2005}:

\begin{equation}
I(X; Y) = \iint P_{X,Y}(x, y) \log \left( \frac{P_{X,Y}(x,y)}{P_X(x) P_Y(y)} \right) dx\,dy,
\label{eMI}
\end{equation}
which quantifies how much knowing $X$ reduces uncertainty about $Y$.\\
Let $S$ be a discrete variable representing the different initial states ($s=0,1,2,3$) and $X\in\mathbb{R}^{70}$ a $70$-dimensional continuous variable representing the values of the homodyne currents at each time step. The joint distribution can be written $P_{X,S}(x, s) = P(s) P(x|s)$, such that Eq.~(\ref{eMI}) becomes
\begin{equation}
I(X; S) = \sum_s P(s) \int P(x|s) \log \left( \frac{P(x|s)}{P(x)} \right) dx.
\label{IG}
\end{equation}
It can also be expressed as the expected Kullback-Leibler (KL) divergence~\cite{Wiley_Sons_2005}:
\begin{equation}
I(S; X) = \mathbb{E}_{s \sim P(s)} \left[ D_{\mathrm{KL}}(P(x|s) \,\|\, P(x)) \right].
\end{equation}
To estimate Eq.~(\ref{IG}), we used a non-parametric approach based on entropy estimates from k-nearest neighbor distances, which was first presented by Kraskov \textit{et al.} in~\cite{Kraskov_2004}. For such numerical estimation, the Python package NPEET-plus (\textit{Non Parametric Entropy Estimation Toolbox}) provides confidence interval estimation for mixed mutual information using bootstrapping.
The currents were randomly sub-sampled to a $3\times10^4$ subset, and (per-feature-)standardized to zero mean and unit variance.
Their dimensionality was then reduced, using Principal Component Analysis, to a single component, which served for the mutual information estimation.

\begin{figure}
    \centering
    \includegraphics[width=0.975\linewidth]{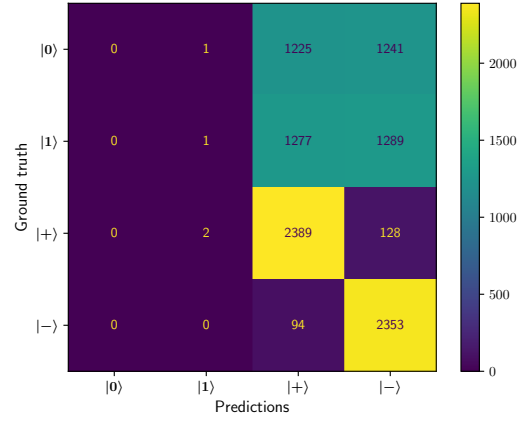}
    \caption{Confusion matrix for our 4-class classification problem. Rows are ground truth while columns are predictions}
    \label{confusion_mat}
\end{figure}

\paragraph{Mutual information between model predictions and ground truth}
Since the predictions and the ground truth are discrete variables, respectively denoted $\hat{y}$ and $y$, Eq.~(\ref{eMI}) becomes
\begin{equation}
    I(\hat{Y}; Y) = \underset{\hat{y},y}{\sum} P(\hat{y},y) \log \left( \frac{P(\hat{y},y)}{P(\hat{y})P(y)} \right),
    \label{discrete_IG}
\end{equation}
which we compute from the confusion matrix of the model, displayed in Fig.~\ref{confusion_mat}.

The confusion matrix $(i,j)$ entries are the empirical joint probabilities $P(\hat{Y}_i, Y_j)$, which we approximate the true probabilities with, and the marginal distributions $P(\hat{Y})$ and $P(Y)$ are the sum of the rows and the columns respectively. Plugging this into Eq.~(\ref{discrete_IG}), we obtain
\begin{equation}
    I(\hat{Y};Y) = 0.1527\ \text{bits}.
\end{equation}

\bibliographystyle{apsrev4-2}
\bibliography{biblio}

\end{document}